# Efficiencies of Aloof-Scattered Electron Beam Excitation of Metal and Graphene Plasmons

Kelvin J. A. Ooi, *Member, IEEE,* Wee Shing Koh, *Senior Member, IEEE,*
Hong Son Chu, Dawn T. H. Tan, and Lay Kee Ang, *Senior Member, IEEE*

***Abstract*—We assessed the efficiencies of surface plasmon excitation by an aloof-scattered electron beam on metals and graphene. Graphene is shown to exhibit high energy transfer efficiencies at very low electron kinetic energy requirements. We show that the exceptional performance of graphene is due to its unique plasmon dispersion, low electronic density and thin-film structure. The potential applications of these aloof-scattered graphene plasmons are discussed in aspects of coherent radiation.**

## I. Introduction

THE interaction of swift electrons with matter provides pathways for evanescent sources of light [1]. One of these pathways involves the excitation of surface plasmons (SPs), which is usually generated by bombardment of swift electrons onto a thin metal film, as has been shown by Ritchie in 1957 [2]. Another method to excite SPs is the aloof-scattering of electrons which was demonstrated by Lecante et al. in 1977 [3], whereby the electron trajectory does not intersect the metal, as schematically shown in Fig. 1. This method involves low-energy transfer and thus minimizes sample damage, in addition to opening up other interesting radiation pathways, such as the plasmon-controlled Cherenkov radiation as has been recently shown by Liu et al. [4].

The aloof-scattering of electrons to excite plasmons on metals is well-studied in the field of electron-loss spectroscopy [1], both in theoretical formulation and experimental observation. In this paper, however, we want to study the aloof-scattering of electrons on graphene, which is still a relatively new area of research with potentially exciting new physics [5].

This work is supported by SUTD (SRG EPD 2011014) and SUTD-MIT IDC grant (IDG21200106 and IDD21200103). L. K. Ang acknowledges the support of a USA AFOSR AOARD grant (14-4020). H. S. Chu acknowledges the support of the National Research Foundation Singapore under its Competitive Research Programme (NRF-CRP 8-2011-07).

K. J. A. Ooi, D. T. H. Tan and L. K. Ang are with the Engineering Product Development, Singapore University of Technology and Design, 20 Dover Drive, 138682 Singapore (e-mail: ricky_ang@sutd.edu.sg).

W. S. Koh and H. S. Chu are with the Electronics and Photonics Department, A*STAR Institute of High Performance Computing, 1 Fusionopolis Way, #16-16 Connexis, Singapore 138632.

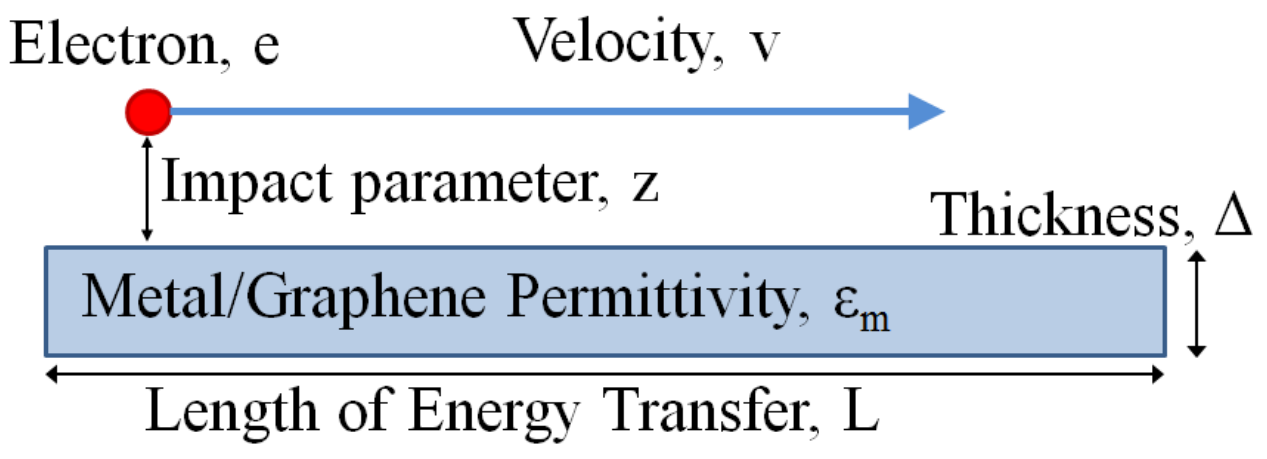

Fig. 1. Aloof-scattering of an electron on a metal/graphene sheet. The electron moves with velocity, v, above the sheet with impact parameter, z. The sheet has a thickness of Δ.

## II. Conditions for Aloof-Scattering Excitation of Surface Plasmons

An aloof-scattered electron can favourably-couple to surface plasmon polaritons (SPPs) of propagation constant given by $k_{sp}=\omega/v$, where ω is the radian frequency and v is the velocity of the electron. As such, the velocity and hence kinetic energy of the electron required to excite SPPs are greatly dictated by the plasmon dispersion of the metal. The plasmon dispersion relation for odd SPP modes on a metal of finite thickness is given by the following equation [6]

$$\coth\left(\sqrt{k_{sp}^2-\omega^2\mu_0\varepsilon_0\varepsilon_m}\,\frac{\Delta}{2}\right)=-\frac{\varepsilon_m}{\varepsilon_d}\cdot\frac{\sqrt{k_{sp}^2-\omega^2\mu_0\varepsilon_0\varepsilon_d}}{\sqrt{k_{sp}^2-\omega^2\mu_0\varepsilon_0\varepsilon_m}} \qquad (1)$$

where $\varepsilon_m=\varepsilon_\infty-\omega_p^2/[\omega(\omega+i\gamma)]$ is the metal permittivity, $\omega_p$ is the plasma frequency, $\varepsilon_\infty$ is the high-frequency dielectric constant of the metal, and γ is the relaxation frequency. $\varepsilon_d$ is the dielectric background permittivity, $\varepsilon_0$ is the free space permittivity, $\mu_0$ is the free space permeability, and Δ is the metal thickness.

### A. *SPP Excitation on Bulk Metals*

For bulk metals, we can approximate the system to be an infinite half-space by taking lim(Δ→∞) in (1). This brings us to the result of the commonly seen dispersion relation

$$k_{sp}=\frac{\omega}{c}\sqrt{\frac{\varepsilon_d\varepsilon_m}{\varepsilon_d+\varepsilon_m}} \qquad (2)$$

where the real part of $k_{sp}$ can be written as

$$\mathrm{Re}(k_{sp}) \approx \frac{\omega}{c}\sqrt{\frac{\varepsilon_d\left(1-x^2\varepsilon_\infty\right)}{\left[1-x^2\left(\varepsilon_d+\varepsilon_\infty\right)\right]}} \tag{3}$$

where $x = \omega/\omega_p$. The plasmon dispersion curve saturates for large $k_{sp}$ at the SP resonance (SPR) frequency when the denominator $\left[1-x^2\left(\varepsilon_d+\varepsilon_\infty\right)\right]$ goes to zero, giving the relation

$$\omega_{sp} = \frac{\omega_p}{\sqrt{\varepsilon_d+\varepsilon_\infty}} \tag{4}$$

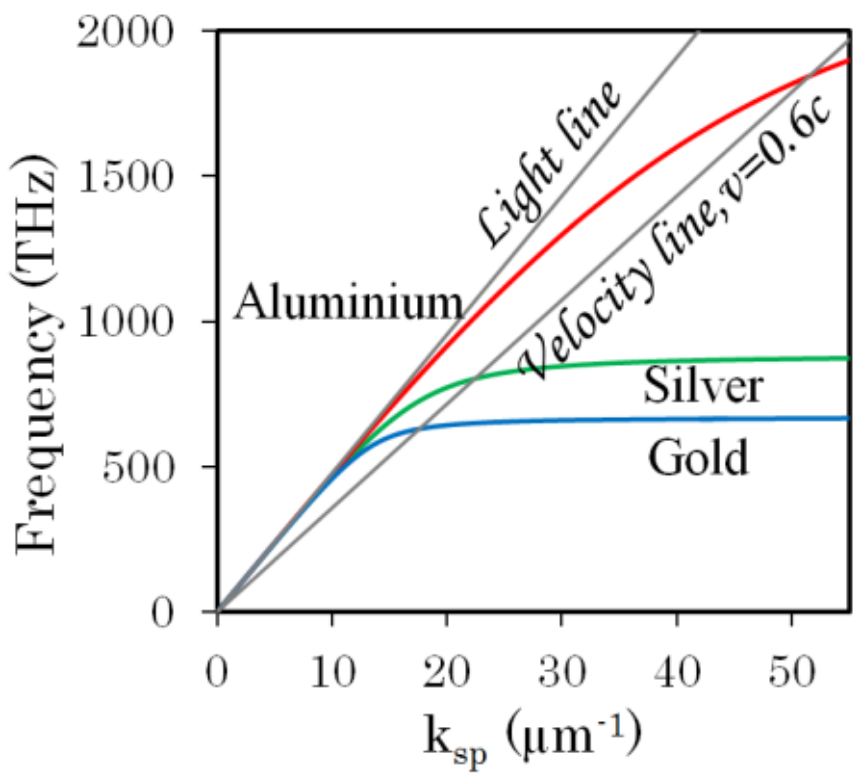

Fig. 2. SPP dispersion curves for bulk aluminium, silver and gold.

At $\varepsilon_d$ and $\varepsilon_\infty$ = 1, we have $\omega_{sp} = \omega_p/\sqrt{2}$, which was the result first observed experimentally by Ritchie [2]. The SPR frequency hence scales with the square-root of the 3-D electronic density, $\omega_{sp} \propto \sqrt{N_{3D}}$.

The excitation of SPPs by aloof-scattering of electrons happens favourably at the frequency where the velocity line intersects the plasmon dispersion line, as shown in Fig. 2. Faster electrons can excite SPPs of lower $k_{sp}$ at the expense of higher electron kinetic energy. While it is possible to excite SPPs of higher $k_{sp}$ using slower electrons, these SPPs would gradually lose their polaritonic nature and become electrostatic SPs. Since the $N_{3D}$ for metals is very high in the order of $10^{28}$ $m^{-3}$, the SPR frequency, and hence the required electron velocity are very high as well. For example, the electron velocity requirements to excite SPs of $k_{sp}$=40μ$m^{-1}$ in vacuum are v=0.4524c and v=0.8378c (where c is the speed of light) for silver and aluminium respectively. These velocities are in the relativistic regime and thus require very high kinetic energies, at 63.2keV and 427keV respectively. Generating SPPs this way is thus not energy-efficient.

### *B. SPP Excitation on Thin Metal Films and Graphene*

Here we want to make the case for doped graphene as an efficient material to excite SPPs by aloof-scattering of electrons. Graphene is an efficient material for two important reasons: firstly, it is a 2-D thin film of vanishing thickness; secondly, graphene's SPR frequency scales with the square-root of the 2-D electronic density as $\omega_{sp} \propto \sqrt{N_{2D}}$, where the $N_{2D}$ is usually in the order of $10^{16}$ $m^{-2}$, which is significantly lower than the $N_{3D}$ for metals [7].

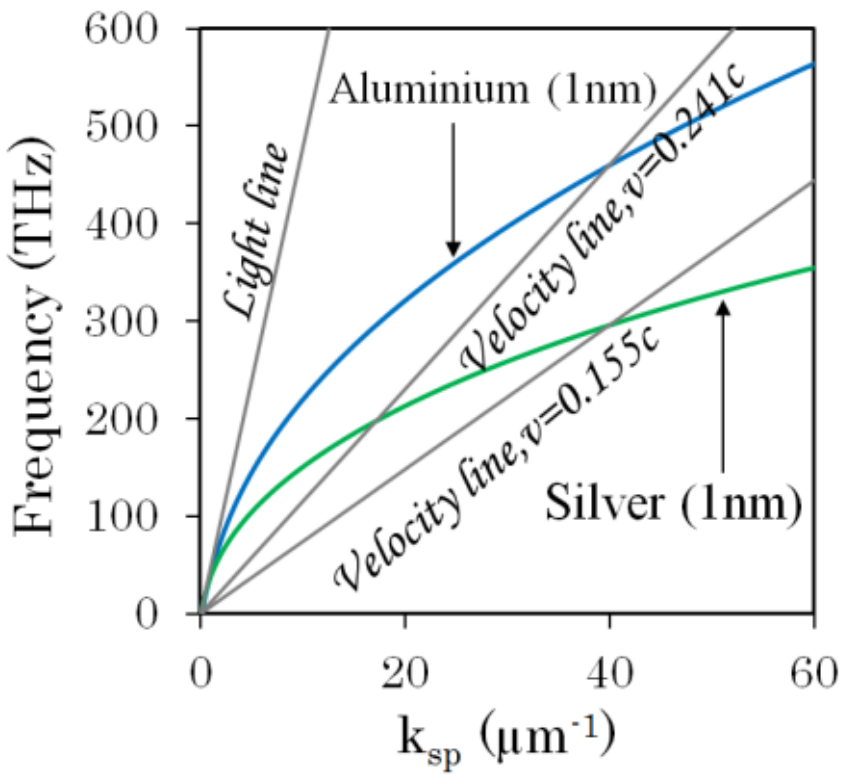

Fig. 3. SPP dispersion curves for 1nm thick aluminium and silver films.

A 2-D thin film of vanishing thickness will have an increased $k_{sp}$. Hypothetically, if the metal films can be made extremely thin (e.g. 1nm thickness), we can take lim(Δ→0) in (1) and thus the dispersion relation becomes

$$k_{sp} = \sqrt{\left(\frac{2\varepsilon_d}{\varepsilon_m\Delta}\right)^2 + \varepsilon_d\left(\frac{\omega}{c}\right)^2} \tag{5}$$

and

$$\mathrm{Re}(k_{sp}) \approx \frac{2\omega^2\varepsilon_d}{\omega_p^2\left[1-x^2\varepsilon_\infty\right]\Delta} \tag{6}$$

As $k_{sp}$ scales inversely with Δ, we can achieve very high $\underline{k_{sp}}$ for low ω and v, as illustrated in Fig. 3.

We further examine the case for graphene, where the plasmon dispersion is given in terms of the 2-D conductivity

$$k_{sp} = \frac{\omega}{c}\sqrt{\varepsilon_d - \left(\frac{2\varepsilon_d\varepsilon_0 c}{\sigma}\right)^2} \tag{7}$$

The optical conductivity, σ is obtained from both the semi-classical model and random-phase approximation, given by

$$\sigma(\omega) = \frac{ie^2E_F}{\pi\hbar^2(\omega+i\gamma)} + \frac{ie^2}{4\pi\hbar}\ln\left(\frac{2|E_F|-(\omega+i\gamma)\hbar}{2|E_F|+(\omega+i\gamma)\hbar}\right) \tag{8}$$

where the first term on the right represents the intraband conductivity and the second term represents the interband conductivity [8]. The constant e is the electronic charge, ħ is the reduced Planck's constant,

$$E_F = \hbar v_F\sqrt{\pi N_{2D}} = \hbar v_F\sqrt{\pi\frac{\varepsilon_d\varepsilon_0}{e}F} \tag{9}$$

is the Fermi level controllable by electric-field F [9], and $v_F = 10^6$m/s is the Fermi velocity of graphene [10]. The graphene conductivity model is valid when $E_F >> k_B T$, where $k_B$ is the Boltzmann constant and T the temperature. It is evident that $k_{sp} \propto 1/\sqrt{N_{2D}}$ and we can thus see a very strongly-deflected plasmon dispersion curve towards the low-frequency mid-infrared regime, as seen from the plot in Fig. 4. Further, we can establish the SPR frequency of graphene when the real part as well as the imaginary part of the conductivity approaches zero, $\sigma'(\omega) \approx \sigma''(\omega) \approx 0$, in which we get

$$\omega_{sp} = \frac{5E_F}{3\hbar} \tag{10}$$

which was the result first obtained by Mikhailov and Ziegler [11], and scales with the $N_{2D}$ like $\omega_{sp} \propto \sqrt{N_{2D}}$. The electron velocities required to excite low-doped graphene plasmons are thus considerably smaller than that for metals, for example, v=0.0178c and v=0.0262c for graphene doped to 0.2eV and 0.4eV respectively, as shown in Fig. 4. These are non-relativistic velocities, thus the electron energies are exceptionally small, at 83.6eV and 175eV respectively.

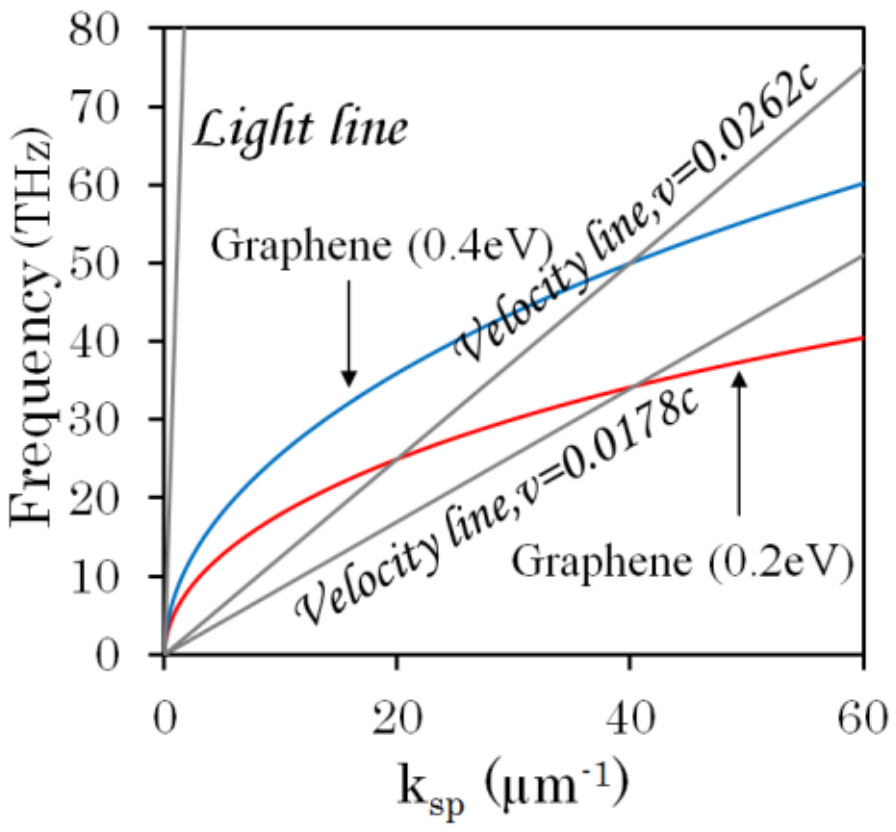

Fig. 4. SPP dispersion curves for graphene doped to 0.2eV and 0.4eV.

## III. Energy Transfer Efficiency

Next, we will discuss the energy transfer efficiency of the aloof-scattered electrons to the graphene plasmons. The energy loss of swift electrons moving with constant velocity in a straight line trajectory can be written as

$$\Delta E = \int \hbar\omega \cdot d\omega \cdot \Gamma(\omega) \tag{11}$$

where $\Gamma(\omega)$ is the loss probability of the electron energy. The loss probability due to graphene plasmons for electron trajectory in the aloof scattering case is given by [1]

$$\Gamma_{SP}(\omega) = \frac{2e^2 L}{\pi\hbar v^2} \cdot K_0\left(\frac{2\omega z}{v\gamma_{LF}}\right) \cdot \mathrm{Im}\left(\frac{r_p}{\varepsilon_0 \varepsilon_d}\right) \tag{12}$$

where L is the length of energy transfer, z is the impact parameter, $\gamma_{LF}$ is the Lorentz contraction factor, and $K_0(x)$ is the zeroth-order modified Bessel function of the second kind. It is immediately observed that $\Gamma_{SP}$ scales inversely with $v^2$, thus the propensity for slower electrons to transfer energy is greater than that for faster electrons. Meanwhile, $K_0(x) \propto -\ln(x)$ for small x and $K_0(x) \propto \exp(-x)$ for large x [12], and thus energy transfer improves either when electrons travel closer to the surface, travel with a non-relativistic velocity, or generate lower $k_{sp}$ SPPs.

Finally, the last term in (12) is the energy loss-function, which represents the propensity of the material to accept the energy transfer, and is given in terms of the p-polarized Fresnel reflection coefficient $r_p$. In general, the energy loss-function for thin metal films and graphene would be very high when compared to bulk metals, because the optical fields could be excited on both sides of the film surface.

### A. *Energy Loss-function of Bulk Metals*

The energy loss-function of bulk metals could be derived as

$$\begin{aligned} \mathrm{Im}\left(\frac{r_p}{\varepsilon_0 \varepsilon_d}\right) &= \mathrm{Im}\left(-\frac{2}{\varepsilon_0 [\varepsilon_d + \varepsilon_m]}\right) \\ &\approx \frac{\omega\gamma}{\varepsilon_0 \omega_p^2} \cdot \frac{2}{\left[1 - x^2(\varepsilon_d + \varepsilon_\infty)\right]^2} \end{aligned} \tag{13}$$

The energy loss-function scales proportionally to ω and γ and inversely proportional to $\omega_p^2$. The energy loss-function will be very large at the SPR frequency defined in (4) when the denominator goes to zero. The large energy acceptance near the SPR frequency, however, would be limited by a small $K_0(x)$ when $k_{sp}$ grows large as well.

### B. *Energy Loss-function of Thin Metal Films*

For thin metal films, $r_p$ is modified to the expression

$$\begin{aligned} r_p &= \frac{\frac{\varepsilon_m - \varepsilon_d}{\varepsilon_d + \varepsilon_m}\left[1 - \exp(-2k_{sp}\Delta)\right]}{1 + \left(\frac{\varepsilon_m - \varepsilon_d}{\varepsilon_d + \varepsilon_m}\right)\left(\frac{\varepsilon_d - \varepsilon_m}{\varepsilon_d + \varepsilon_m}\right)\exp(-2k_{sp}\Delta)} \\ &\approx 1 + \frac{2k_{sp}\Delta}{\exp(2k_{sp}\Delta) - 2k_{sp}\Delta - 1} \end{aligned} \tag{14}$$

For $k_{sp} \cdot \Delta << 1$, $r_p$ is a very large number. Thus, the energy loss-function for thin film metals will be very large below the SPR frequency. Near the SPR frequency, however, (14) will be reduced to $r_p$ for bulk metals.

### C. *Energy Loss-function of Graphene*

We can write $r_p$ for graphene as [5]

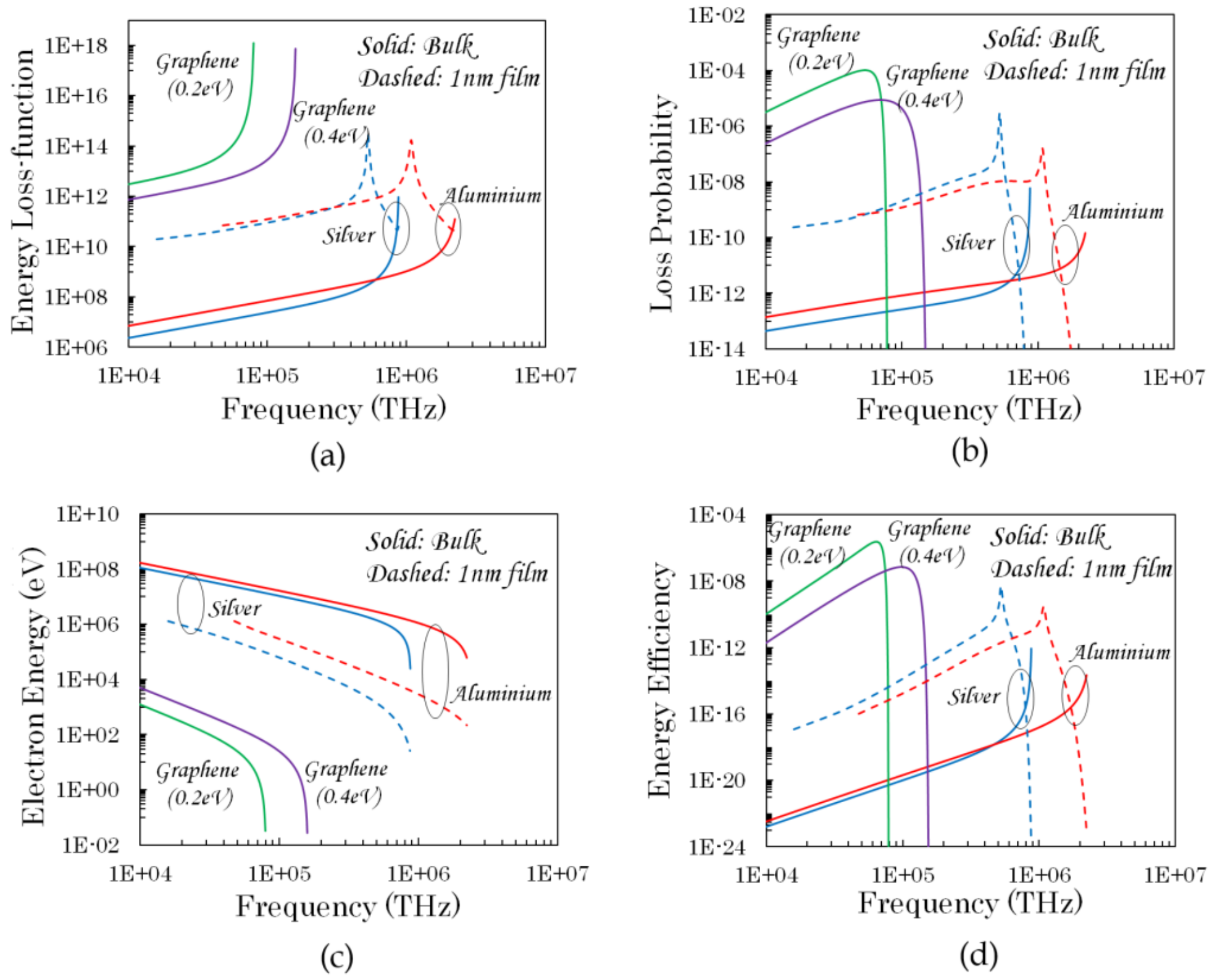

Fig. 5. Energy transfer performance comparison for 6 types of material: bulk aluminium, bulk silver, 1nm aluminium film, 1nm silver film, and graphene doped to 0.2eV and 0.4eV. Relevant parameters for comparison are shown in the following logarithmic plots: (a) energy loss-function, (b) loss probability, (c) electron kinetic energy, and (d) the energy transfer efficiency.

$$r_p = 1 - \frac{2\varepsilon_{d1}}{\varepsilon_{d1} + \varepsilon_{d2} + \dfrac{i\sigma k_{sp}}{\varepsilon_0 \omega}} \tag{15}$$

where $\varepsilon_{d1}$ is the incident background permittivity and $\varepsilon_{d2}$ is the substrate permittivity. Below the SPR limit, the energy loss-function for graphene is derived to be

$$\mathrm{Im}\left(\frac{r_p}{\varepsilon_0 \varepsilon_d}\right) \approx \left[\frac{\pi \hbar^2 c}{e^2}\right]^2 \cdot \frac{16\varepsilon_0 \omega \gamma}{E_F^2} \tag{16}$$

which scales proportionally to ω and γ and inversely to $E_F^2$ like the case for bulk metals.

### *D. Energy Transfer Performance Comparison*

In Fig. 5 we show some relevant parameters for the performance comparison of the energy transfer efficiency. The materials used for the comparison are aluminium and silver (material parameters taken from ref. [13,14]), and graphene (γ assumed to be $2\times10^{12}$Hz). In Fig. 5(a) and 5(b), the energy loss-function and loss probability are plotted using the expression given in (12). The energy transfer length in the calculations are normalized to 1m. In Fig. 5(c), the electron kinetic energy at the condition for favorable coupling is obtained using the usual relativistic kinetic energy equation

$$KE = m_e c^2 \left[\frac{1}{\sqrt{1-(v/c)^2}} - 1\right] \tag{17}$$

Finally, in Fig. 5(d), the energy transfer efficiency is obtained from the ratio of the product of loss probability with unit plasmon energy, ħω, over the electron kinetic energy.

Several features stood out in the figures. Graphene's energy loss-function is exceptionally high owing to its low $N_{2D}$ values, and grows even larger near the SPR frequency. However, the graphene's loss probability near the SPR frequency is limited by the the $K_0(x)$ term in (12) due to large $k_{sp}$ values. Meanwhile, thin metal films have overall better energy transfer performance compared to their bulk counterparts, as has been predicted from (14), except that the performance deteriorates near the SPR frequency when the large $k_{sp}$ values becomes the limiting factor, similar to graphene. Overall, from Fig. 5(d), the energy transfer efficiency for graphene is on average 4-orders higher than thin metal films, and 10-orders higher than bulk metals.

In Fig. 6 we looked at some electric-field maps of the PIC-simulated electron-excited plasmons using VORPAL [15]. To facilitate comparison, we selected 3 material systems with corresponding electron speeds that excite a characteristic $k_{sp}$ of ~40μm$^{-1}$: (a) bulk silver and electron beam speed of 0.45c, (b) 1nm silver film and electron beam speed of 0.155c, and (c) 0.2eV doped graphene and electron beam speed of 0.0178c. The electric-field maps capture the moment when the electron has traversed 400–500nm along the material surface. It is observed that even with the vastly-varied electron beam speeds (and hence the electron kinetic energies), the excited plasmon electric-fields are in the same order of magnitude, which clearly corroborates with the high excitation energy efficiency of graphene plasmons compared to silver plasmons.

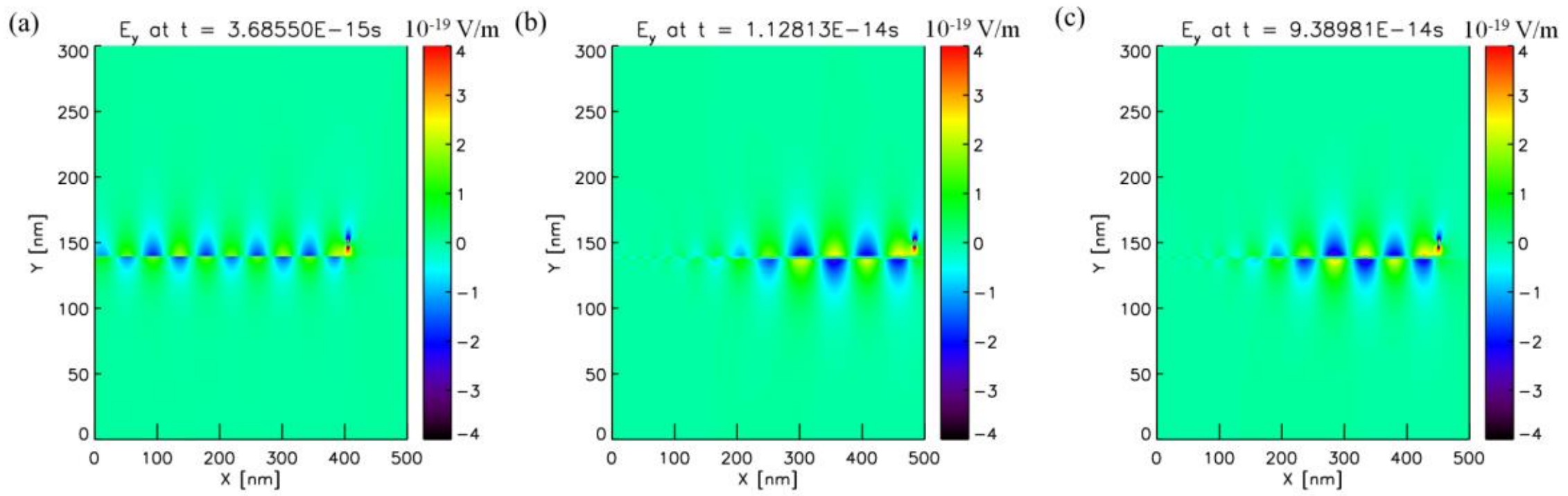

Fig. 6. Electric-field maps for aloof-scattered electron excitation of plasmons for (a) bulk silver using electron speed of 0.45c, (b) 1nm silver film using electron speed of 0.155c, and (c) 0.2eV doped graphene using electron speed of 0.0178c. In all cases, the excited characteristic $k_{sp} \sim 40\mu m^{-1}$.

## IV. Potential Applications of Aloof-Scattered Graphene Plasmons

Graphene's high energy transfer efficiency could be used for applications in generating coherent radiation. For example, one of the possible applications is the plasmon-controlled Cherenkov radiation as mentioned previously. Liu et al. have shown that the radiation spectrum of Cherenkov radiation, which is usually broadband, could be made frequency-selective by first exciting the metal plasmon modes, and the radiation frequency can be tuned over a narrow spectral range in the visible and UV regime by tuning the electron beam velocity [4]. We took the configuration of Liu et al. and replaced the metal layer with graphene. Fig. 7 shows that for a 100keV electron beam, we could potentially excite Cherenkov radiation modes in the 1–2 THz frequency spectrum. Thus, graphene could potentially be used for efficient THz radiation sources. Moreover, the graphene Cherenkov radiation spectrum has increased tunability: we can tune both the electron beam velocity and also the Fermi level of graphene through an electrostatic-gating.

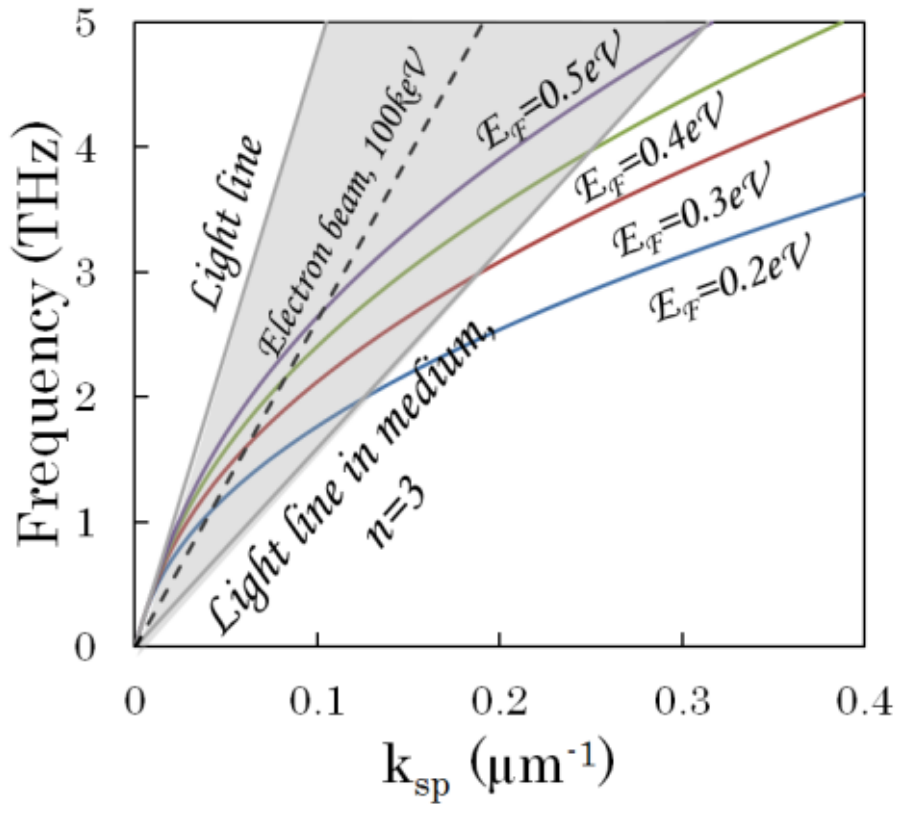

Fig. 7. Cherenkov radiation is possible in the shaded cone bounded by the light line in vacuum and in a medium having refractive-index of n=3. An electron beam of 100keV kinetic energy could excite graphene plasmons in the THz spectrum.

In cases where low energy electron beams are preferred and Cherenkov-type of radiation is not possible, it is still possible to induce coherent radiation from gratings [16]. Liu et al. has recently shown that the coherent radiation generated from graphene-on-gratings are 400 times stronger than conventional dielectric materials.

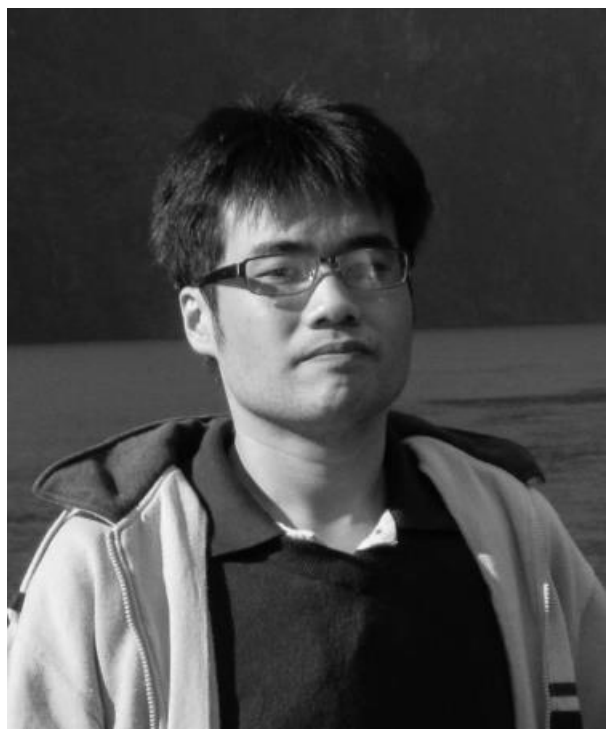

**Kelvin J. A. Ooi (GSM'11–M'15)** received his B.Eng. degree (Hons.) and Ph.D. degree in electrical and electronic engineering (EEE), Nanyang Technological University (NTU), Singapore, in 2010 and 2014 respectively, both majoring in photonics.

He joined the Singapore University of Technology and Design (SUTD) late 2013 as a Research Assistant, and is currently a Postdoctoral Research Fellow under the joint supervision of Prof. Lay Kee Ang and Prof. Dawn T. H. Tan. His current research interests include nanophotonics, electrically-excited plasmonics, graphene plasmonics, and nonlinear optics.

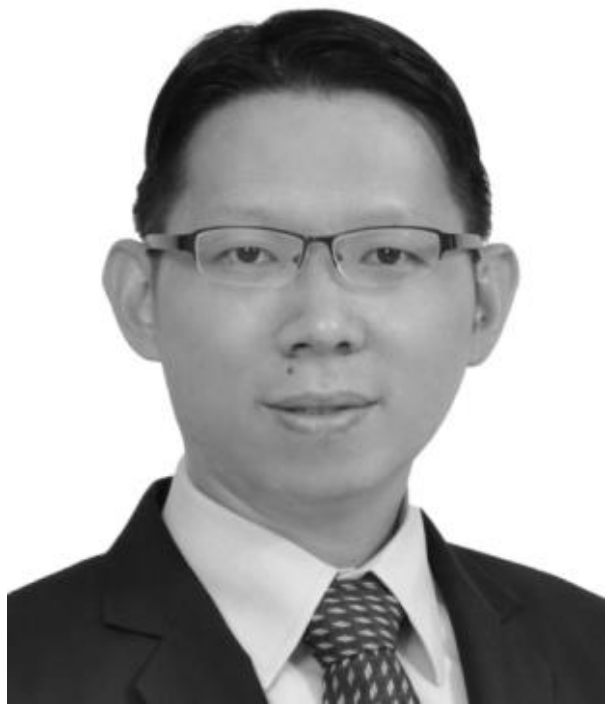

**Wee Shing Koh (S'03–M'07–SM'14)** was born in Singapore in 1977. He received his B.Eng. (Hons.) and Ph.D. degrees in electrical and electronics engineering (EEE) from the Nanyang Technological University (NTU), Singapore, in 2002 and 2007, respectively.

He joined the A*STAR Institute of High Performance Computing (IHPC), Singapore, as a Research Engineer since June 2006. He was the inaugural IHPC Independent Investigatorship awardee from 2011-2013 and is currently a Capability Group Manager and Scientist at the Electronics & Photonics Department in IHPC. His research interests include organic solar cell devices, computational plasmonics for thin film solar cells, sensors and microelectronics, field emission theory, multi-dimensional space-charge-limited transport, and particle-in-cell simulations for plasma, nanophotonic and nanoelectronic applications. He has published more than 50 journal and conference papers.

Dr. Koh is the current chairman of the IEEE NPSS chapter in Singapore.

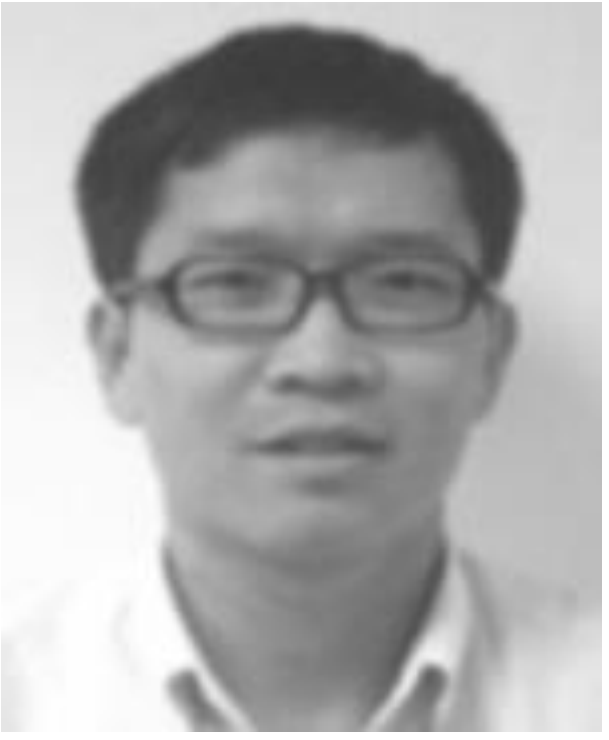

**Hong Son Chu** received a B.E. degree in electronics and telecommunications engineering from the Ho Chi Minh City University of Technology, Vietnam, in 1999. He obtained a M.S. degree in propagation, telecommunications and remote sensing in June 2000, and a Ph.D. degree in electronics in June 2004, both from the University of Nice-Sophia Antipolis, France.

From Nov. 2004 to Nov. 2005, he joined the Computational Electromagnetics Research Laboratory (CERL), University of Victoria, Victoria, Canada as an NSERC Post-Doctoral Fellow. Since 2006, he has been with the A*STAR Institute of High Performance Computing (IHPC), Singapore, as a Scientist and Capability Group Manager of the Photonics and Plasmonics group under the Electronics and Photonics Department. His research interests are in plasmonic and photonic nanostructures from fundamentals to devices and applications, and computational techniques for electromagnetics from microwave to optical wavelengths. He has authored/co-authored over 50 technical papers in international referred journals and conferences.

**Dawn T. H. Tan** obtained her Ph.D. and M.S. from the University of California at San Diego and her B.A.Sc from the University of British Columbia.

She is an Assistant Professor at the Singapore University of Technology and Design (SUTD), Singapore. From 2013 to 2014, she was a visiting scholar at the Massachusetts Institute of Technology (MIT), Cambridge, MA, USA. She was previously a senior engineer at Luxtera – a leader in CMOS photonics, where she worked in the field of silicon photonics. She has published three book chapters and over 40 refereed journal and conference articles in the field of integrated optics, silicon photonics and nonlinear optics.

Prof. Tan was the recipient of the Best Paper Award at the Workshop on Information Optics, 2009.

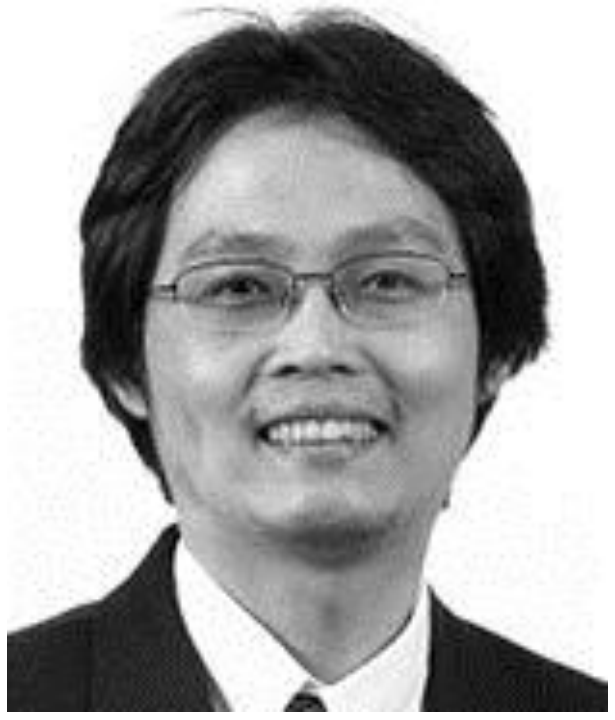

**Lay Kee Ang (GSM'95–M'00–SM'08)** received the B.S. degree from the Department of Nuclear Engineering, National Tsing Hua University, Hsinchu, Taiwan, in 1994, and the M.S. and Ph.D. degrees from the Department of Nuclear Engineering and Radiological Sciences, University of Michigan, Ann Arbor, MI, USA, in 1996 and 1999, respectively.

He was awarded a fellowship to work as a Los Alamos National Laboratory Director Postdoctoral Fellow in the Plasma Physics Applications Group in the Applied Physics Division from 1999 to 2001. He was an Assistant Professor and a tenured Associate Professor in the Division of Microelectronics, School of Electrical and Electronic Engineering, Nanyang Technological University, Singapore, from 2001 till 2011. He has been with the Singapore University of Technology and Design, Singapore, since 2011, and is currently the Ph.D. Program Director and a faculty member of the Engineering Product Development (EPD) pillar. His research interests include electron emission from novel materials, ultrafast laser induced photocathode, space charge limited current, multipactor discharge, plasmonics, and charge injection into solids. He has published more than 70 journal papers on these topics.

Prof. Ang was the founding chairman of the IEEE NPSS chapter in Singapore in 2012. He was awarded several Window of Science Awards from AFOSR-AOARD to be short term visiting scientist to USA.